\documentclass[aps,12pt]{revtex4-1}
\usepackage{amsmath}
\usepackage{graphicx}% Include figure files
\usepackage{dcolumn}% Align table columns on decimal point
\usepackage{bm}% bold math
\usepackage{epsfig}
\usepackage{graphics}
\begin{document}

%\preprint{}

\title{A metric for gravitational collapse of dust around a Schwarzschild black hole}
%\hspace{20pt} \\
%\maketitle
\author{Recai Erdem}
\email{recaierdem@iyte.edu.tr}
\author{Bet{\"{u}}l Demirkaya}
\email{betuldemirkaya@iyte.edu.tr}
%\author{Kemal G{\"u}ltekin}
%\email{kemalgultekin@iyte.edu.tr}
\affiliation{Department of Physics \\
{\.{I}}zmir Institute of Technology\\
G{\"{u}}lbah{\c{c}}e, Urla 35430, {\.{I}}zmir, Turkey}

\date{\today}

\begin{abstract}
We consider the problem of gravitational collapse of a fluid under the effect of a Schwarzschild black hole (e.g. a primordial one) that suddenly forms inside the fluid. We assume the fluid initially be a uniform dust. Starting from this configuration we obtain a class of metrics under some assumptions. We find that the metric we obtain includes the dust collapse as a subcase. After discussing some basic properties of the solution, we discuss the case of dust collapse in more detail. We find that the radial and tangential pressures outside the horizon may take positive or negative values depending on the values of the parameters.
\end{abstract}
%\pacs{}

%\keywords{}

\maketitle

\section{introduction}
Gravitational collapse
is the main engine behind the formation of stars, galaxies, clusters of galaxies out of small initial inhomogeneities. Therefore, gravitational collapse, in particular
the gravitational collapse of a fluid around a black hole,  attracted attention since the early days of the theory of general relativity.
The early major studies on this topic are done by McVittie in 1933 \cite{McVittie}, Tolman in 1934 \cite{Tolman}, Oppenheimer and Snyder in 1939 \cite{Oppenheimer}. Later in 1947 Bondi \cite{Bondi}
introduced a metric (called Lemaitre-Tolman-Bondi metric) that may be converted to many other metrics (including those of Oppenheimer and Snyder, Robertson-Walker, Schwarzschild)  after specifying
some particular values and some coordinate transformations while McVittie's metric is not a subcase of this metric \cite{generalized-Bondi}. The common feature of all these metrics (including McVittie's)
is that all these correspond to non-accrediting black holes \cite{generalized-Bondi,Kaloper}. In 1951 Vaidya \cite{Vaidya} introduced a wholly different metric in the sense that its
(Schwarzschild-like) black hole accumulates or radiates massless particles, so it is accrediting. All the major metrics for gravitational collapse at present essentially belong to one of
McVittie, Lemaitre-Tolman-Bondi, or Vaidya metrics or their variations. In the following paragraphs, we derive a new metric that describes an accrediting black hole - fluid system that initially consists of
a Schwarzschild black hole and an
homogeneous and isotropic dust that consists of particles of some mass $m$.

{\section{framework}

\subsection{Preliminaries}

The most general spherically symmetric metric can be expressed as \cite{Joshi,Bambi}
\begin{equation}
ds^2=-e^{2\lambda}dt^2+e^{2\psi}dr^2+R^2(d\theta^2+\sin^2\theta\, d\phi^2), \label{eq:1}
\end{equation}
where $\lambda=\lambda(r,t)$, $\psi=\psi(r,t)$, $R=R(r,t)$. The corresponding non-vanishing components of Einstein tensor are
\begin{eqnarray}
&& G^0_{\;0}\,=-\,\frac{F^\prime}{R^2 R^\prime}\,+\,\frac{2\dot{R}\,e^{-2\lambda}}{R\,R^\prime}\left(\dot{R} ^\prime-\dot R \lambda^\prime-\dot \psi R^\prime\right), \label{eq:1a} \\
&& G^1_{\;1}\,=-\,\frac{\dot F}{R^2\dot R }\,-\,\frac{2R^\prime\,e^{-2\psi}}{R\,\dot{R}}\left(\dot{R}^\prime-\dot R \lambda^\prime-\dot \psi R^\prime\right), \label{eq:1b} \\
&& G^0_{\;1}\,=-\,e^{2\psi-2\lambda}\,G^1_{\;0}\,=\,
\frac{2e^{-2\lambda}}{R}\left(\dot{R} ^\prime-\dot R \lambda^\prime-\dot \psi R^\prime\right), \label{eq:1c} \\
&& G^2_{\;2}\,=\,G^3_{\;3}\,=\,\frac{e^{-2\psi}}{R}\,\left[\left(\lambda^{\prime\prime}+{\lambda^\prime}^2- \lambda^\prime\psi^\prime\right)R\,+\,R^{\prime\prime}+ R^\prime\lambda^\prime-R^\prime\psi^\prime\right] \nonumber\\
&&-\,
\frac{e^{-2\lambda}}{R}\,\left[\left(\ddot{\psi}+\dot{\psi}^2- \dot{\lambda}\dot{\psi}\right)R\,+\,\ddot{R}+ \dot{R}\dot{\psi}-\dot{R}\dot{\lambda}\right],
\label{eq:1d}
\end{eqnarray}
where the indices $0,1,2,3$ denote $t,r,\theta,\phi$, and
\begin{equation}
F(r,t)=R(1-e^{-2\psi}R'^2+e^{-2\lambda}\dot R^2) \label{eq:4}
\end{equation}
is the Misner-Sharp mass function.

\subsection{Derivation of a specific Misner-Sharp mass function}

If black hole is not accrediting i.e. if $G^0_{\;1}\,=\,0$ (e.g. as in Lema{\^i}tre-Tolman-Bondi metric), then we may take the energy momentum tensor $T^\mu_\nu$ in
Einstein equations to be of type I in the Hawking-Ellis classification \cite{types} i.e. as
\begin{equation}
T^\mu_\nu=\text{diag}(-\rho,P_1,P_2,P_3) \label{eq:2}
\end{equation}
in comoving coordinates. The corresponding Einstein equations for (\ref{eq:1}) result in
\begin{equation}
\frac{F^{'}}{R^2 R^{'}}\,=\, 8\pi\rho, \label{eq:3a}
\end{equation}
\begin{equation}
\frac{\dot F}{R^2\dot R }\,=\,-8\pi P_r, \label{eq:3b}
\end{equation}
\begin{equation}
\dot R ^{'}-\dot R \lambda^{'}-\dot \psi R^{'}\,=\,0, \label{eq:3c}
\end{equation}
\begin{eqnarray}
&&\frac{e^{-2\psi}}{R}\,\left[\left(\lambda^{\prime\prime}+{\lambda^\prime}^2- \lambda^\prime\psi^\prime\right)R\,+\,R^{\prime\prime}+ R^\prime\lambda^\prime-R^\prime\psi^\prime\right] \nonumber\\
&&-\,
\frac{e^{-2\lambda}}{R}\,\left[\left(\ddot{\psi}+\dot{\psi}^2- \dot{\lambda}\dot{\psi}\right)R\,+\,\ddot{R}+ \dot{R}\dot{\psi}-\dot{R}\dot{\lambda}\right]\,=\,-8\pi P_\theta,
\label{eq:3d}
\end{eqnarray}
where primes and overdots denote the partial derivative with respect to $r$ and $t$, respectively, $P_r=P_1$, $P_\theta=P_2=P_3$.

Now we impose the initial condition to specify the Misner-Sharp function, namely, the gravitational collapse is induced by a black hole that suddenly forms inside dust.
To this end, first consider (\ref{eq:3a})
\begin{equation}
\frac{F^{'}}{R^2 R^{'}}= 8\pi\rho \ \Longrightarrow \ F=\int_0^{r} F^{'}d\tilde{r}=8\pi\int_0^{r}\rho R^2 R^{'}d\tilde{r} \label{eq:5}
\end{equation}
\begin{equation}
R'=\frac{\partial R}{\partial r} \ \Longrightarrow \ R^{'}dr=dR \mid_{t=\text{constant}}=dR_c,
\end{equation}
where $R_c=R(r,t_c)$ with $t_c=\text{constant}$.
Hence, (\ref{eq:5}) may be also expressed as
\begin{equation}
F=8\pi\int_0^{R_0}\rho\tilde{R}_c^2 d\tilde{R}_c. \label{eq:6}
\end{equation}
In other words, effectively we integrate over $\tilde{R}_c=R_c(\tilde{r},t_c)$ in (\ref{eq:5}).

Note that the infinitesimal volume element for (\ref{eq:1}) \cite{Misner} is
\begin{equation}
d^3V=4\pi R^2 e^\psi dr \label{eq:7}.
\end{equation}
The mechanical energy of a test particle of mass m, excluding its potential energy, is
\begin{equation}
E=m\,e^\lambda \frac{dt}{ds} \label{eq:8}.
\end{equation}
This expression may be obtained by dividing both sides of (\ref{eq:1}) by $ds^2$, and then identifying the result as the local Minkowski expression $m^2=E^2-\vec{p}^2$ for unit mass.
Note that this is the relevant quantity in an energy-momentum tensor rather than the total mechanical energy of the particle $m\,e^{2\lambda} \frac{dt}{ds}$ (that may be obtained from $\frac{\partial\,L}{\partial\,t^\prime}$, where $L=-m\frac{\sqrt{-ds^2}}{dt}$, $t^\prime=\frac{dt}{ds}$). The relevance of this identification may be better seen by considering the non-relativistic limit of Einstein equations for weak fields, namely, Possion equation, $\vec{\nabla}^2\phi=4\pi\,G\,\sum\,m_i\delta(\vec{r}-\vec{r}_i)$ for a set of point masses. It is evident from the Poisson equation that the source term (i.e. the energy-momentum tensor term) does not contain the gravitational potential, the potential term is on the left hand side of the equation (i.e. in the Einstein tensor part).
If we assume that the fluid is made of particles of mass m and number density $n$, then the energy density of the fluid is
\begin{equation}
\rho=En, \label{eq:9}
\end{equation}
where $E$ is given by (\ref{eq:8}). After using (\ref{eq:7}), (\ref{eq:8}), (\ref{eq:9}); (\ref{eq:5}) becomes
\begin{equation}
F=2\int_0^r R^{'}e^{-\psi}\rho\, d^3V=2\int_0^r R^{'}e^{-\psi}E\,n\, d^3V \label{eq:10}.
\end{equation}
In the case (e.g. for Schwarzschild metric)
\begin{equation}
R^{'}\frac{dt}{ds}e^{\lambda-\psi}=1 \label{eq:11}.
\end{equation}
(\ref{eq:10}), after the use of (\ref{eq:8}) and (\ref{eq:11}), becomes
\begin{equation}
F=2m\int_0^r n\, d^3V \label{eq:12},
\end{equation}
which is the twice of the total rest mass energy inside a sphere of radius $r$. In the following we will assume that Eqs. (\ref{eq:11}) and (\ref{eq:12}) are satisfied.

We consider a model where the universe initially consists of a homogeneous isotropic dust that is made of particles of mass m. We assume that a Schwarzschild black hole of mass $M$ suddenly appears in this universe.
This may be considered to model a primordial black hole that suddenly forms (e.g due to quantum fluctuations) at some time $t_i$ in the presence of a uniform dust \cite{PBH-matter-dominated} of initial number
density $n_0$.

Just before the formation of the black hole, the metric is that of the Robertson-Walker metric (that we assume to be spatially flat), so we have
\begin{equation}
    e^{2\lambda^{(b)}_{0}}=1, \ \ e^{2\psi^{(b)}_{0}}=a^2(t), \ \ R^{(b)^{2}}=a^2(t)r^2.  \label{eq:14a}
\end{equation}
At initial times after the formation of the black hole, locally the situation is almost equivalent to that of a single black hole since the dust initially does not have a local gravitational field (except its cosmological
effect) because of the homogeniety and the isotropy of the dust. Therefore, initially just after the formation of the black hole we have
\begin{equation}
  e^{2\lambda^{(a)}_{0}}\simeq (1-\frac{2M}{r_0}), \ \    e^{2\psi^{(a)}_{0}}\simeq (1-\frac{2M}{r_0})^{-1}\;, \ \
  R^{(a)^{2}}\simeq\, r_0^2, \label{eq:14b}
\end{equation}
provided that the cosmological effects may be neglected. As is evident from (\ref{eq:14b}), $r_0$ corresponds to $R$ after the collapse if the effect of cosmological expansion is neglected. Its detailed form will be derived in the next section.

The total number of particles inside a comoving radial distance $r$ due to collapse of the dust may be expressed as
\begin{equation}
N=\int_0^r n\, d^3V=\int_0^{r} n{_0}\, d^3 V{_0}, \label{eq:13}
\end{equation}
where $n=\frac{n_0}{a^3}$ with $n_0$ being a constant for non-interacting cosmological fluids while it may depend on the spatial coordinates in general, and $d^3V=a^3d^3V_0$.

Eq. (\ref{eq:13}) for (\ref{eq:14a}) and (\ref{eq:14b}) gives the number of particles (in a sphere of radius $r$) $N^{(b)}$ before and $N^{(a)}$ after the formation of the black hole as
\begin{equation}
N^{(b)}=\int_0^{r}n^{b}d^3V^{b}=\int_0^{r}n_0^{b}d^3V_0^{b}=\frac{4\pi}{3}n_0^{b}r^{3}\label{eq:15aa},
\end{equation}
where $n_0^{(b)}$ is a constant, and
\begin{equation}
N^{(a)}=\int_0^{r}n_0^{a}d^3V_0^{a} \label{eq:15ab},
\end{equation}
with
\begin{equation}
d^3V_0^{b}=4\pi\, r^2dr\;,~~~~
d^3V_0^{a}=4\pi(R^{(a)})^{2}e^{\psi^{(a)}_{0}}dr_0=4\pi r_0^2 \big(1-\frac{2M}{r_0}\big)^{-1/2}dr_0\;, \label{eq:15b}
\end{equation}
where the superscripts $^b$ and $^a$ refer to the time just before and after the introduction of the black hole.

Because the mass function just before and after the formation of the black hole must coincide at the initial time $t=t_i$, Eq.(\ref{eq:12}) implies that we must set $N^{(b)}|_{t=t_i}\,=\,N^{(a)}|_{t=t_i}$. This may be insured if we let
\begin{equation}
N^{(a)}=\int_0^{r}n_0^{a}d^3V_0^{a}=\frac{4\pi}{3}n_0\,r_0^{3} \label{eq:15ba}
\end{equation}
where $n_0=n_0^{b}$, and
$r_0=r$ at $t\,\leq\,t_i$, and $r_0=r_0(r,t)$ is a function of $r$ and $t$ for $t\,>\,t_i$. Therefore, we make the following plausible assumption
for $t\,>\,t_i$
\begin{equation}
	F\,=\,F\left(r_0\right)\,=\,2M\,+\,\frac{8\pi}{3}m\,n_0\,r_0^3,
	\label{eq:16}
\end{equation}
where the contribution due to the black hole is included by the term $2M$. The explicit form of $r_0$ (provided that Eqs. (\ref{eq:11}) and (\ref{eq:16}) are satisfied) will be derived in the next section.
%It may be easily checked that $F$ satisfies the regularity conditions derived in \cite{Joshi2}, namely,
%$\frac{F_{,r}}{R^\prime}\,<\,\infty$, $\frac{\left(F_{,r}\right)_{R=r}}{r^2}\,<\,\infty$, $\frac{\left(F_{,R}\right)_{R=r}}{r^2}\,<\,\infty$ at $r=0$ (where %the subscript $_,$ denotes partial derivative) that must be satisfied
%at initial times.
%The additional regularity condition $\left[p_2\right]_{r=0}\,<\,\infty$ at initial time should be checked after specifying the metric. We will find that $p_2$ %is wholly due to accretion, and it turns out to be infinite at $r=0$ for dust after we introduce accretion. In fact, this also turns to be the case for energy %density and $p_1$ at $r=0$. In fact this is not unexpected since $r=0$ is a singularity of the black hole.

A consistency check for (\ref{eq:16}) may be done as follows. Before the emergence of the black hole we have a dust. Eq.(\ref{eq:4}) in this case may be rearranged as \cite{Bambi,Joshi3,Joshi5}
\begin{equation}
\dot{R}^2\,=\,\frac{F}{R}\,+\,s,
\label{eq:16aa}
\end{equation}
where $s$ is a constant that should be determined by initial conditions. Initially $M=0$, and we may take $r_0=r$,
so (\ref{eq:16aa}) (after the use of (\ref{eq:16}) for $M=0$) reduces to
\begin{equation}
\frac{\dot{a}^2}{a^2}\,=\,\frac{8\pi}{3}m\,\frac{n_0}{a^3}\,\,+\,\frac{s}{a^2(t)\,r^2}\,=\,\frac{8\pi}{3}\rho\,\,+\,\frac{s}{a^2(t)\,r^2}.
\label{eq:16aaa}
\end{equation}
This together with Friedmann equation for zero spatial curvature implies $s=0$. It, in turn, implies that the dust is marginally bound \cite{book,Joshi4} i.e. the dust particles initially have zero peculiar velocity. In other words,
taking $F$  initially having contribution only due to rest mass energies of the particles is consistent. Just after the emergence of the black hole, $F$ picks up an additional term 2M, hence becomes $2M\,+\,\frac{8\pi}{3}m\,n_0\,r_0^3$.
Eq.(\ref{eq:10}) and the condition (\ref{eq:11}) insure that this form of $F$ is conserved during the collapse.

After these remarks, we return to our main discussion. For a mass function that depends only on $r_0$ (as the one given (\ref{eq:16}) we have
\begin{eqnarray}
	F^{'}=\frac{\partial F}{\partial r}= \frac{\partial  r_0}{\partial r}\frac{\partial F}{\partial r_0}~, ~~ \dot F= \frac{\partial F}{\partial t}= \frac{\partial r_0}{\partial t}\frac{\partial F}{\partial r_0}.
\end{eqnarray}
We identify $r_0(r,t)$ by the physical radial position of the particles where the effect of cosmological expansion is subtracted i.e.
\begin{equation}
	R\left(r,t\right)=a\left(t\right)r_0\left(r,t\right).
	\label{eq:20}
\end{equation}
Note that $r_0(r,t)$ before the formation of the black hole is identical with the radial coordinate $r$ so it is independent of time while after the formation of the black hole it has time dependence due to
local gravitational collapse that will be derived in the next section (namely, Eq.(\ref{eq:29})).
Hence, (\ref{eq:3a})  and (\ref{eq:3b}) become
\begin{equation}
\frac{F^{'}}{R^2R^{'}}= \frac{r_0^{'} \frac{\partial F}{\partial r_0}}{a^3\left(t\right)r_0^2r_0^{'}}= \frac{1}{a^3r_0^2}\frac{\partial F}{\partial r_0}= 8 \pi \rho,
\label{eq:21a}	
\end{equation}
\begin{equation}
\frac{\dot F}{R^2\dot R }= \frac{\dot r_0 \frac{\partial F}{\partial r_0}}{a^2\left(t\right)r_0^2\dot R }= \frac{1}{a^3r_0^2}\frac{\partial F}{\partial r_0}-\frac{\dot a}{a^3r_0\dot R}\frac{\partial F}{\partial r_0}
= \frac{1}{a^3r_0^2}\left(1-\frac{\dot a r_0}{\dot R }\right)\frac{\partial F}{\partial r_0}  = -8 \pi P.
\label{eq:21b}	
\end{equation}
Here
\begin{equation}
	\rho=\frac{M}{4\pi r_0^2}\delta\left(r_0\right) + \tilde{\rho}\left(r_0,t\right),
	\label{eq:22}
\end{equation}
where the term with the delta function in (\ref{eq:22}) stands for the black hole and $\tilde{\rho}$ is the energy density of the fluid.
Note that, in the case $\dot a=0$ (i.e in the case where cosmological expansion is neglected, Eq.(\ref{eq:21a}) and Eq.(\ref{eq:21b}) give an equation of state parameter -1
(as for a cosmological constant). On the other hand, in the case $\dot r_0=0$, i.e. in the case where there is no gravitational collapse due to the black hole we have $\dot{R}= \dot{a}\, r_0$, so $P=0$ in
Eq.(\ref{eq:21b}), i.e the fluid behaves as dust.

\section{a relevant metric}

In this section first we will show that there is no metric that simultaneously satisfies (\ref{eq:3c}), (\ref{eq:11}), (\ref{eq:16}) and (\ref{eq:21a}), (\ref{eq:21b}). Then, in order to keep the interesting implications of
(\ref{eq:21a}), (\ref{eq:21b}) that follows from (\ref{eq:16}), we will relax the condition (\ref{eq:3c}).
Finally, we find a solution with a non-vanishing $G^0_{\;1}$ component that approaches zero at large distances from the black hole.
In other words, the derived metric approximately satisfies (\ref{eq:21a}), (\ref{eq:21b}) with (\ref{eq:16}) at large distances from the black hole.
Therefore, the main conclusion obtained at the end of the preceeding section remains the same.

\subsection{An unsuccessful attempt}

After taking the derivative of (\ref{eq:11})  with respect to time and dividing by itself, we obtain
\begin{equation}
\dot R ^\prime + \dot \lambda R^\prime-\dot{\psi} R^\prime=0 \;\Rightarrow~ \dot{R}^\prime - \dot{\psi} R^\prime = - \dot{\lambda} R^\prime .
\label{eq:24}
\end{equation}
We impose (\ref{eq:24}) together with (\ref{eq:4}), (\ref{eq:3c}) and (\ref{eq:16}). After substituting (\ref{eq:24}) in (\ref{eq:3c}), we obtain
\begin{equation}
 \dot \lambda R^\prime+ \lambda^\prime \dot{R}=0 \Rightarrow \frac{\lambda^\prime}{\dot \lambda}= - \frac{R^\prime}{\dot R }.
\label{eq:25}
\end{equation}
Note that
\begin{eqnarray}
 \frac{\lambda^{'}}{\dot \lambda}=\frac{\big(\frac{\partial e^{2\lambda}}{\partial r}\big) }{\big(\frac{\partial e^{2\lambda}}{\partial t}\big)}~, ~~\frac{R^{'}}{\dot R }= \frac{\big(\frac{\partial R^2}{\partial r}\big)}{\big(\frac{\partial R^2}{\partial t}\big)}.
\end{eqnarray}
One may check that Eq.(\ref{eq:25}) has the following solution,
\begin{eqnarray}
	e^{2\lambda}= \alpha_1 + \alpha_2 \frac{f\left(t\right)}{g\left(r\right)}~, ~~ R^2=\alpha_3 + \alpha_4\left[f\left(t\right)g\left(r\right)\right]^\beta
		\label{eq:26},
\end{eqnarray}
where $\alpha_1,\alpha_2,\alpha_3,\alpha_4, \beta$ are arbitrary constants; and $f\left(t\right),g\left(r\right)$ are arbitrary functions of $t$ and $r$, respectively.
Initially we have both Schwarzschild black hole and homogeneous dust i.e. $e^{2\lambda_0}= 1 -\frac{2(M+\frac{4}{3}mn_0r^3)}{r}$ suggests that one may take
\begin{eqnarray}
\alpha_1=1,~ && \alpha_2=-2~,~~ g\left(r\right)=\left(\frac{M+\frac{4}{3}mn_0r^3}{r}\right)^{-1}.
	\label{eq:27}
\end{eqnarray}
Then $R^2\left(r,t\right)$ may be specified by $\alpha_3=0$, $\alpha_4=1$, $\beta=2$ such that
\begin{equation}
R\left(r,t\right)\,=\,r\,f(t).
\label{eq:28}
\end{equation}

After substituting (\ref{eq:26}) (with (\ref{eq:27}) and (\ref{eq:28})) in (\ref{eq:24}), we get
\begin{equation}
\dot{\psi}\,=\,\frac{\dot{f}}{f}-\frac{(M+\frac{4\pi}{3}mn_0r^3)\dot{f}}{r-2(M+\frac{4\pi}{3}mn_0r^3)\,f},
\label{eq:28b}
\end{equation}
which, after integration, results in
\begin{equation}
e^{2\psi}\,=\,Q(r)\,f^2\left(r-2(M+\frac{4}{3}mn_0r^3)\,f\right).
\label{eq:28b}
\end{equation}
where $Q(r)$ is an arbitrary function of $r$ (whose form to be determined by initial conditions).

The $\lambda$, $\psi$, $R$ given in (\ref{eq:26}) and (\ref{eq:28b}) may be substituted in Eq.(\ref{eq:4}) to obtain the corresponding F as
\begin{equation}
F(r,t)=r\,f(t)\left[1-\frac{1}{Q(r)\left(r-2(M+\frac{4}{3}mn_0r^3)f(t)\right)}+\frac{r^3\dot{f}(t)^2}{\left(r-2(M+\frac{4}{3}mn_0r^3)f(t)\right)} \right].
\end{equation}
We find that this F does not have the form given in Eq.(\ref{eq:16}). This implies that the conditions (\ref{eq:11}) and (\ref{eq:3c}) are not consistent. Therefore, we must drop either of these conditions. In the following subsection, we drop the condition (\ref{eq:3c}) and seek a solution.

\subsection{Accretion as a cure}

We have found above that (\ref{eq:11}) and (\ref{eq:3c}) are not consistent. We have to drop one of them.
We choose to drop the condition (\ref{eq:3c}) since (\ref{eq:11}) leads to the attractive possibility of a local dark energy effect given in (\ref{eq:21a}) and (\ref{eq:21b}).
Therefore, now we are left with the equation (\ref{eq:24}) together with (\ref{eq:4}) and (\ref{eq:16}), so $R$, $\psi$, $\lambda$ remain under-determined.
Moreover, the violation of the condition (\ref{eq:3c}) (that allows a non-zero $G^0_{\;1}$) must be small to insure approximate validity of (\ref{eq:21a}) and (\ref{eq:21b}). One may try to keep one of $R$ or $g_{00}$ obtained in (\ref{eq:26}) intact and determine the other only by using (\ref{eq:24}) (supplemented with (\ref{eq:4}) and (\ref{eq:16})) to find the corresponding solution with the hope of obtaining a $G^0_{\;1}$ whose smallness may be controlled by some parameter. Because the form of the derivatives of $R$ in both of (\ref{eq:24}) and (\ref{eq:3c}) are the same while this is not the case for the derivatives of $\lambda$ and $\psi$, it is safer to keep $R$ as the one given in (\ref{eq:28}) and determine $e^{2\lambda}$ and $e^{2\psi}$ by using (\ref{eq:24}),(\ref{eq:4}) and (\ref{eq:16}).

Eq.(\ref{eq:24}), after using (\ref{eq:28}), may be reduced to
\begin{equation}
\frac{\dot{f}}{f}\,=\,\frac{\partial\,\left(\psi-\lambda\right)}{\partial\,t},
\label{eq:28c}
\end{equation}
which may be integrated to get
\begin{equation}
e^{-2\lambda}\,=\,Q(r)\,f^2\,e^{-2\psi}\, , \label{eq:28d}
\end{equation}
where $Q$ is an arbitrary function of $r$ (whose form to be determined by initial conditions). Eq.(\ref{eq:4}) may be rewritten as
\begin{equation}
e^{2\psi}\,=\,\left[1+e^{-2\lambda}\dot{R}^2-\frac{\,F(r)}{R}\right]^{-1}R^{\prime\;2}.
\label{eq:28e}
\end{equation}
Eq.(\ref{eq:28e}) may be solved for $e^{2\psi}$ after using (\ref{eq:28d}),  (\ref{eq:16}) and (\ref{eq:24}) (i.e. $R=f\,r$) as
\begin{equation}
e^{2\psi}\,=\,\frac{f^2\left(1-r^2Q\,\dot{f}^2\right)}{\left[1-\frac{\,F(r_0)}{R}\right]},
\label{eq:28f}
\end{equation}
where $F(r_0)$ is given by (\ref{eq:16}). Then one may substitute (\ref{eq:28f}) in (\ref{eq:28d}) to obtain
\begin{equation}
e^{2\lambda}\,=\,\frac{\left(1-r^2Q\,\dot{f}^2\right)}{Q\left(1-\frac{\,F(r_0)}{R}\right)}.
\label{eq:28g}
\end{equation}

Because we have dropped the condition (\ref{eq:3c}), we expect a non-vanishing $G^0_{\;1}$ in general. After using a Mathematica code, we find that this is really the case. Therefore,
we must
add a non-vanishing $T^0_{\;1}$ term to the energy-momentum tensor in Einstein field equations. This may be done by changing the type of the energy momentum tensor from type I to type II \cite{types} (i.e. one may find a frame where the energy-momentum tensor in a local Minkowski frame may be transformed into the following form by local Lorentz transformations)
\begin{eqnarray}
\left(T^{\mu\nu}\right)\,=\,\left(\begin{array}{cccc}
\xi+\sigma&\sigma&0&0\\
\sigma&-\xi+\sigma&0&0\\
0&0&p_2&0\\
0&0&0&p_3\end{array}\right),
\label{eq:28r}
\end{eqnarray}
where $\xi$, $\sigma$, $p_2$, $p_3$ are some functions of the coordinates.
After applying an arbitrary  Lorentz transformation in $x_0-x_1$ plane to (\ref{eq:28r}) and then  comparing  the result of an explicit calculation of the components of the corresponding
Einstein tensor for  (\ref{eq:24}), (\ref{eq:28f}) and (\ref{eq:28g}) by using a Mathematica code shows that (\ref{eq:28r}) is really the relevant energy-momentum tensor.

It is evident that before the formation of the black hole, there will be no density contrast, so we must impose $\lim_{M\to0} f\left(t_i\right)=a(t_i)$.
\begin{equation}
R\left(r,t\right)\,=\,r\,f(t)\,=\,a(t)\,\left[\frac{f(t)}{a(t)}\right]r\,.
\label{eq:28a}
\end{equation}

Comparison of (\ref{eq:28}) with (\ref{eq:20}) gives
\begin{equation}
r_0\,=\,h(t)\,r, \label{eq:29}
\end{equation}
where
\begin{equation}
h(t)=\frac{f(t)}{a(t)}. \label{eq:30}
\end{equation}

Hence, after using (\ref{eq:29}), (\ref{eq:30}) and (\ref{eq:16}) (namely $F\left(r_0\right)=2M+2mn_0\frac{4\pi}{3}r_0^3$) in (\ref{eq:28f})and (\ref{eq:28g}) we find the relevant metric as
\begin{eqnarray}
ds^2&=&-\frac{\left(1-r^2Q(r)\,\dot{f}^2(t)\right)}{Q(r)\,\left(1-\frac{2\,\left(M+mn_0\frac{4\pi}{3}\left(\frac{f(t)}{a(t)}\right)^3r^3\right)}{f(t)\,r}\right)}
dt^2 \nonumber \,+\,\frac{f^2(t)\,\left(1-r^2Q(r)\,\dot{f}^2(t)\,\right)}{\left(1-\frac{2\,\left(M+mn_0\frac{4\pi}{3}\left(\frac{f(t)}{a(t)}\right)^3r^3)\right)}{f(t)\,r}\right)}
dr^2
\nonumber \\
&&
\,+\,f^2(t)\,r^2(d\theta^2+\sin^2\theta d\phi^2), \label{eq:31a}
\end{eqnarray}
where $f(t)~\rightarrow\,a(t)$ and $M\rightarrow\,0$ as $t\rightarrow\,t_i$. Different choices of $f(t)$ are expected to give different evolutions of the system.

The conditions that (\ref{eq:31a}) must reduce to the Schwarzschild metric if the dust is removed (i.e. when $n_0=0$) while it should reduce to the Robertson-Walker metric for the initial homogeneous dust (i.e. for $M\,=\,0$) specifies $Q(r)$ as
\begin{equation}
Q(r)\,=\,\frac{1}{\left(1-\frac{2M}{r}\right)^2}.
\label{eq:31aa}
\end{equation}
When we set $n_0=0$ in (\ref{eq:31a}) with (\ref{eq:31aa}) we automatically obtain the Schwarzschild metric. On the other hand, (\ref{eq:31a}) with (\ref{eq:31aa}) reduces to the spatially flat Robertson-Walker metric after we set $M\,=\,0$ and use $f(t)\,\rightarrow\,a(t)$. Moreover if we also impose that the fluid is a homogeneous isotropic dust then we should impose $\frac{\dot{a}^2}{a^2}\,=\,\frac{8\pi}{3}\rho\,=\,\frac{8\pi}{3}m\frac{n_0}{a^3}$ (i.e. $a(t)\propto\,t^\frac{2}{3}$) to make the corresponding pressures become zero as we will see in the next section. This implies that the metric in (\ref{eq:31a}) is not limited to description of gravitational collapse of dust although it is obtained by considering the gravitational collapse of a fluid that is initially dust. The solution (\ref{eq:31a}) describes a family of solutions where the metric for gravitational collapse of dust is a subcase case of (\ref{eq:31a}).

\section{some general implications of the metric}

\subsection{General behaviour of the solution}

One may find the components of Einstein tensor corresponding to (\ref{eq:31a}) by using a computer code e.g. Mathematica. Although the expressions are rather complicated, still one may get some general information about the $ G^0_{\;0}$ and the  $G^1_{\;1}$ components of Einstein tensor as follows.
Eqs. (\ref{eq:1a}), (\ref{eq:1b}), (\ref{eq:1c}) may be expressed as
\begin{eqnarray}
&& G^0_{\;0}\,=-\,\frac{F^\prime}{R^2 R^\prime}\,+\,\Delta\,G^0_{\;0} \,=-\,\frac{F^\prime}{R^2 R^\prime}\,+\,\frac{\dot{R}}{R^\prime}\,\,G^0_{\;1}, \label{eq:1aa} \\
&& G^1_{\;1}\,=-\,\frac{\dot F}{R^2\dot R }\,+ \,\Delta\,G^1_{\;1}\,=-\,\frac{\dot F}{R^2\dot R }\,- \,\frac{R^\prime}{\dot{R}}e^{2\lambda-2\psi}\,G^0_{\;1},\label{eq:1ba} \\
&& G^0_{\;1}\,=-\,e^{2\psi-2\lambda}\,G^1_{\;0}\,=\,
\frac{2e^{-2\lambda}}{R}\left(\dot{R} ^\prime-\dot R \lambda^\prime-\dot \psi R^\prime\right). \label{eq:1ca}
\end{eqnarray}
It is evident from the above expressions that the additional contributions to $G^0_{\;0}$ and $G^1_{\;1}$  when compared to those in Eqs.(\ref{eq:21a}) and (\ref{eq:21b}), namely, $\Delta\,G^0_{\;0}$ and $\Delta\,G^1_{\;1}$ are determined by the values of $G^0_{\;1}$, $\frac{\dot{R}}{R^\prime}$, $\frac{R^\prime}{\dot{R}}e^{2\lambda-2\psi}$.
From  (\ref{eq:28}) and (\ref{eq:31a}) we find
\begin{eqnarray}
\frac{\dot{R}}{R^\prime}\,=\,\frac{r\dot{f}(t)}{f(t)}~,~~~ \frac{R^\prime}{\dot{R}}e^{2\lambda-2\psi}\,=\,\frac{1}{r\,Q(r)\,f(t)\dot{f}(t)}\,=\,
\frac{\left(1-\frac{2\,M}{r}\right)^2}{r\,f(t)\dot{f}(t)}.
\label{eq:1caa}
\end{eqnarray}
Therefore, the signs of $\Delta\,G^0_{\;0}$ and $\Delta\,G^1_{\;1}$ are essentially determined by the sign of $\dot{f}$ and $G^0_{\;1}$.

%Note that
%\begin{equation}
%\frac{\Delta\,G^1_{\;1}}{\Delta\,G^0_{\;0}}\,=\,-\left(\frac{R^\prime}{\dot{R}}\right)^2e^{2\lambda-2\psi}\,=\,-\frac{1}{Q(r)\dot{f}^2 r^2}
%\label{eq:1caaa}
%\end{equation}

 By using Mathematics we find the components of Einstein tensor for (\ref{eq:31a}) and (\ref{eq:31aa}) in the limit of $r\,\rightarrow\,\infty$ as
 \begin{eqnarray}
 &&\mbox{As}~\; r\,\rightarrow\,\infty \nonumber \\
 && G^0_{\;0}\,\rightarrow\,-\frac{8 \pi\,m\,n_0 \{f(t) \left[3 \dot{a}(t) \dot{f}(t)+2 a(t) \ddot{f}(t)\right]+a(t) \dot{f}^2(t)\}}{3 a(t)^4 \dot{f}^2(t)}                                    \label{i1a} \\
 && G^0_{\;1}\,\rightarrow\, 0    \label{i1b} \\
 && G^1_{\;1}\,\rightarrow\,
 \frac{8 \pi\,m\,n_0 \{f(t) \dot{a}(t)-a(t) \dot{f}(t)\}}{a(t)^4 \dot{f}(t)}
 \label{i1c} \\
 && G^2_{\;2}\,=\,G^3_{\;3}\,\rightarrow\,
 \frac{4}{3} \pi\,\frac{m\,n_0}{a^3(t)} \{-\frac{3f^2(t) \ddot{a}(t)}{a(t) \dot{f}^2(t)}+\frac{3f^2(t) \dot{a}^2(t)}{a^2(t) \dot{f}^2(t)}-\frac{3f(t) \dot{a}(t)}{a(t) \dot{f}(t)} \nonumber \\
&& -\frac{2f^{(3)}(t) f^2(t)}{\dot{f}^3(t)}+\frac{2f^2(t) \ddot{f}^2(t)}{ \dot{f}^4(t)}-\frac{4f(t) \ddot{f}(t)}{\dot{f}^2(t)}-2\}
 \label{i1d}
 \end{eqnarray}

In the limit of $r\,\rightarrow\,\infty$ the only non-vanishing component of the Einstein tensor must be $G^0_{\;0}$) if we assume that the fluid is initially dust since at an infinite distance from the black hole the black hole can not have any effect. In fact, in the limit $M\,\rightarrow\,0$ we obtain the same result for the components of Einstein tensor as in the above equations. On the other hand, if we also let
$f(t)\,=\,a(t)\,\propto\,t^\frac{2}{3}$ these equations reduce to
\begin{eqnarray}
&&\mbox{As}~\;r\,\rightarrow\,\infty\,,~~ f(t)\,\rightarrow\,a(t)\,\propto\,t^\frac{2}{3} \nonumber \\
&&G^0_{\;0}\,\rightarrow\,-\frac{8 \pi\,m\,n_0}{a(t)^3}~,~\;G^0_{\;1}\,=\, G^1_{\;1}\,=\,
 G^2_{\;2}\,=\,G^3_{\;3}\,\rightarrow\,0   \label{i2}
 \end{eqnarray}
 i.e. the fluid reduces to usual uniform dust as expected. However, if we only require
 $f(t)\,=\,a(t)$ (in the limit of $r\,\rightarrow\,\infty$) without specifying $a(t)\,\propto\,t^\frac{2}{3}$ then we find that $G^2_{\;2}\,=\,G^3_{\;3}$ is non-vanishing and $G^0_{\;0}$ does not have the evolution of a free dust. This implies that the metric obtained in (\ref{eq:31a}) is more general than that of dust evolution in the presence of a black hole although we have obtained it by starting from a dust. In other words, (\ref{eq:31a}) describes the evolution of a cosmological fluid (that includes dust as a special case) in the presence of a black hole.

It is evident from (\ref{i1b}) that the second parts of (\ref{eq:1aa}), (\ref{eq:1ba}) become negligible at large $r$.
Hence we obtain equations similar to (\ref{eq:21a}) and (\ref{eq:21b}) for large values of $r$, namely,
\begin{equation}
\frac{\frac{\partial F}{\partial r_0}}{a^3\left(t\right)r_0^2}=\,\frac{a(t)}{2\pi\,f^4(t)r^4}M\,\delta(r)+\frac{8\pi\,m\,n_0}{a^3(t)}\,=\, 8 \pi \rho^{(1)}\,\simeq\, 8 \pi \rho,
\label{eq:32a}	
\end{equation}
\begin{eqnarray}
\frac{1}{a^3r_0^2}\left(1-\frac{\dot a r_0}{\dot R }\right)\frac{\partial F}{\partial r_0}
&=&\left[1-\left(\frac{\dot{a}(t)}{a(t)}\right)\left(\frac{f(t)}{\dot{f}(t)}\right)\right]\left(\frac{a(t)\,M}{2\pi\,f^4(t)r^4}\delta(r)+\frac{8\pi\,m\,n_0}{a^3}\right) \nonumber \\
&=& -8 \pi P_r^{(1)}\,\simeq\, -8 \pi P_r,
\label{eq:32b}	
\end{eqnarray}
where the upper scripts $^{(1)}$ refers to the contribution due to the first part of the Einstein equations, and
we have substituted the explicit forms $R$ and $r_0$ and included a step function in $F$ to obtain the delta function in $\rho$ i.e. in (\ref{eq:32a}) and (\ref{eq:32b}) we have used
\begin{equation}
F\left(r_0\right)=2M\Theta(\vec{r}_0)+2mn_0\frac{4\pi}{3}r_0^3.  \label{eq:32}
\end{equation}.

 Eqs. (\ref{eq:32a}) and (\ref{eq:32b}) suggest that the conclusions obtained after (\ref{eq:21a}) and (\ref{eq:21b}) remain the same at large distances from the black hole:
 The overall fluid, in general, has two extremes well-known extremes, the cosmological dust for $f(t)\,=a\,(t)$ (i.e. for $M\rightarrow\,0$ at initial time $t_i$)
 and the local cosmological-constant-like behaviour for $|\frac{\dot{a}}{a}|\,\ll\,|\frac{\dot{f}}{f}|$ (i.e. in the absence of a cosmological expansion or contraction).
To see the implications of (\ref{eq:32a}) and (\ref{eq:32b}) more clearly, it will be useful to identify the terms responsible for local gravitational collapse and the cosmological expansion
 more clearly. $a(t)$ is taken to be the scale factor for the cosmological expansion. (\ref{eq:20}) combined with (\ref{eq:29})
 implies that $r_0$ corresponds to $R$ after the effect of cosmological expansion is subtracted, and $h(t)$ specifies the local gravitational collapse. The equation of state parameter may be determined from (\ref{eq:32a}) and (\ref{eq:32b}) as
 \begin{equation}
\omega\,=\,\frac{P}{\rho}\,=\,-\left[1-\left(\frac{\dot{a}(t)}{a(t)}\right)\left(\frac{f(t)}{\dot{f}(t)}\right)\right]\,=\,-\frac{1}{1\,+\,\frac{\dot{a}(t)}{a(t)}\frac{h(t)}{\dot{h}(t)}}.
\label{eq:33}	
\end{equation}
 Eq. (\ref{eq:33}) implies that $\omega$ depends both on $\frac{\dot{a}}{a}$ and $\frac{\dot{h}}{h}$. We observe that, when $|\frac{\dot{a}}{a}|\,\ll\,|\frac{\dot{h}}{h}|$ (i.e. when the
 cosmological expansion is negligible with respect to the local contraction due to the local gravitational collapse) the equation of state of the system approaches the equation of state of cosmological
 constant while, for $|\frac{\dot{a}}{a}|\,\gg\,|\frac{\dot{h}}{h}|$ the equation of state of the system approaches that of dust. In the phenomenologically relevant case of cosmological expansion
 (i.e. $\frac{\dot{a}}{a}\,>\,0$) and the local gravitational collapse  (i.e. $\frac{\dot{h}}{h}\,<\,0$), the equation of state parameter $\omega$ may take all possible values between $-\infty\,<\,\omega\,<\,+\infty$.
 In the cases where the universe is contracting both at cosmological and at local scales or expanding both at cosmological and at local scales  we would have
 $-1<\,\omega\,<\,\,0$.

\subsection{A brief overview of the apparent horizons and the singularities of the solution}

We must address the essential points related to the causal structure of the spacetime described by the metric obtained in the last section in order to have a clearer
picture of the evolution of the corresponding black hole - fluid system. To this end we first obtain the apparent horizon(s) of the metric. The apparent horizon(s) may be determined by the condition $k^\mu_{\;;\mu}\,=\,0$
where $k^\mu$ stand for the tangent vectors of affinely parameterized radial null geodesics \cite{book,galaxies}. In the case of spherical symmetry (as is the case for (\ref{eq:31a})) one may also obtain the apparent horizon(s)
through either of the conditions $g^{\mu\nu}\left(\partial_\mu\,R\right)\left(\partial_\nu\,R\right)\,=\,0$ or by $g^{RR}\,=\,0$ after converting the coordinates of (\ref{eq:31a}) (i.e. the comoving gauge) to the Kodama
gauge \cite{Faraoni2}. In either of these equations one obtains the following equation for the apparent horizon(s)
\begin{equation}
\frac{8\pi}{3}\,m\,n_0\left(\frac{f(t)}{a(t)}\right)^3\,r^3\,-\,f(t)\,r\,+\,2M\,=\,0.
\label{eq:40}	
\end{equation}
The roots of (\ref{eq:40}) are \cite{galaxies}
\begin{eqnarray}
&&r_1\,=\,\frac{1}{\sqrt{2\pi\,m\,n_0}}\left(\frac{a(t)\sqrt{a(t)}}{f(t)}\right)\,\sin{\psi}
 \label{eq:41a}	\\
 &&r_2\,=\,\sqrt{\frac{3}{8\pi\,m\,n_0}}\left(\frac{a(t)\sqrt{a(t)}}{f(t)}\right)\,cos{\psi}\,-\,\frac{1}{\sqrt{2\pi\,m\,n_0}}\left(\frac{a(t)\sqrt{a(t)}}{f(t)}\right)\,\sin{\psi}
 \label{eq:41b}	\\
 &&r_3\,=\,-\sqrt{\frac{3}{8\pi\,m\,n_0}}\left(\frac{a(t)\sqrt{a(t)}}{f(t)}\right)\,cos{\psi}\,-\,\frac{1}{\sqrt{2\pi\,m\,n_0}}\left(\frac{a(t)\sqrt{a(t)}}{f(t)}\right)\,\sin{\psi}
 \label{eq:41c}	
\end{eqnarray}
with $\sin{3\psi}\,=\,\frac{1}{a(t)\sqrt{a(t)}}\,3M\,\sqrt{8\pi\,m\,n_0}$. $r_3$ is not physically relevant since a radial coordinate can not be negative. $r_1$ coincides with the apparent horizon of
the black hole - fluid system. This may be easily seen in the limit of the energy density of the fluid not being much larger than the present day energy density of the universe.
In that case $3M\,\sqrt{8\pi\,m\,n_0}\,\ll\,1$ i.e.
$sin3\psi\,\simeq\,3\psi\,\ll\,1$
since $\sqrt{8\pi\,m\,n_0}$ is some multiple of $H_0\,\sim\,\frac{1}{\left(3\times\,10^8\,meters/second\right)\times\,3\times\,10^{17}\,seconds}$ while $M$ is in the order of the Schwarzschild event horizon that is
at the order of kilometers for
stellar black holes and smaller than $\sim\,10^{14}\,kilometers$ for known supermassive black holes.  This, in turn, implies that initially $r_1\,\simeq\,2M$ for such a case which is the Schwarzschild event horizon of the black hole.
In a similar way we obtain $r_2\,\sim\,\frac{1}{H_0}$ i.e. the cosmological horizon
for the initial time (in the same limit as $r_1$).

In fact, the above formula for the apparent horizons is essentially the same as the one for the Schwarzschild-de-Sitter-Kottler metric given by \cite{galaxies}
 \begin{eqnarray}
ds^2&=&-\left(1-\frac{2M}{R}-H^2R^2\right)dT^2 \,+\,\frac{dR^2}{\left(1-\frac{2M}{R}-H^2R^2\right)}
\,+\,R^2(d\theta^2+\sin^2\theta d\phi^2), \label{eq:42}
\end{eqnarray}
 where $H=\sqrt{\frac{\Lambda}{3}}$ with $\Lambda\,>\,0$ being a positive cosmological constant.
The close similarity between the apparent horizon structures of (\ref{eq:42}) and (\ref{eq:31a}) is not accidental. (\ref{eq:1}) may be transformed into a form similar to (\ref{eq:42}) (see Appendix A), namely,
 \begin{eqnarray}
ds^2&=&-e^{2\lambda}\,\left(1-\frac{U^2}{\Gamma^2}\right)\,W^2\,dT^2 \,+\,\frac{dR^2}{\Gamma^2-U^2}\,dR^2
\,+\,R^2(d\theta^2+\sin^2\theta d\phi^2), \label{eq:43}
\end{eqnarray}
where
\begin{equation}
U\,=\,e^{-\lambda}\dot{R}~~,~~\Gamma\,=\,e^{-\psi}\,R^\prime~~,~~dT\,=\,\frac{1}{W}\left(dt+\beta\,dR\right)
\label{eq:44}
\end{equation}
with $W=W(t,R)$ satisfying $\frac{\partial\frac{1}{W}}{\partial\,R}=\frac{\partial\frac{\beta}{W}}{\partial\,t}$ and
\begin{equation}
\beta\,=\,\frac{e^{2\psi}\dot{R}}{R^{\prime\;2}\left(e^{2\lambda}-e^{2\psi}\frac{\dot{R}^2}{R^{\prime\;2}}\right)}.
\label{eq:45}
\end{equation}
In the case of (\ref{eq:31a}) we obtain
\begin{equation}
\Gamma^2\,-\,U^2\,=\,e^{-2\lambda}\dot{R}^2\,-\,e^{-2\psi}\,R^{\prime\;2}\,=\,1-\frac{2M+mn_0\frac{4\pi}{3}\left(\frac{f(t)}{a(t)}\right)^3r^3}{r\,f(t)}.
\label{eq:46}
\end{equation}
We see that the $g^{RR}$ terms of both of (\ref{eq:42}) and (\ref{eq:43}) for (\ref{eq:31a}) have the same form. This is the reason for the identical forms of the apparent horizons for (\ref{eq:42}) and (\ref{eq:31a}).
However, (\ref{eq:42}) and (\ref{eq:31a})  are still different since the coefficients of the $dT^2$ terms in (\ref{eq:42}) and (\ref{eq:43}) for (\ref{eq:31a}) are not the same and they cannot be set to be the same.
This may be seen as follows. First we equate the coefficients of $dT^2$ terms in (\ref{eq:42}) and (\ref{eq:43}) for (\ref{eq:31a}). Then, we solve the equation for $W$. Next, we express $W$ in terms of $R$ and $t$ by replacing $r$ by $\frac{R}{f(t)}$.
Finally, we find $\frac{\partial\frac{1}{W}}{\partial\,R}-\frac{\partial\frac{\beta}{W}}{\partial\,t}$ (by using Mathematica) which we observe to be non-vanishing in general.

Next, we discuss the singularity structure for this metric. In studies of gravitational collapse of a fluid, determining whether the singularities induced by gravitational collapse are naked or hidden behind a horizon is
an important issue. To this end
one may re-express $F(r,t)=R(1-e^{-2\psi}R'^2+e^{-2\lambda}\dot R^2)$ as $\dot{R}^2\,=\,\frac{F}{R}\,+\,s$ where $s$ is a function of $r$ and $t$
in general (while in the case of dust it may be taken to be a constant), and determine the behaviour of $\frac{F}{R}\,+\,s$ that is expressed in terms of $R=r\,f(t)$ and $t$ in a region of space-time \cite{Joshi3,Joshi4,Joshi5}.
The resulting equation
may be used to find the time $t_f\,-\,t_i$ for evolution of a spherical shell of radius $R_i=r\,f(t_i)$ to a spherical shell of radius $R_f=r\,f(t_f)$. If initially there is no singularity as in the case of dust collapse, then
$t_f\,-\,t_i$ may be used to find the time $\Delta\,t_{AH}$ for formation of the apparent horizon by letting $t_f=t_{AH}$ (the time when the apparent horizon reaches the coordinate $r$) while it may be used to find the time
$\Delta\,t_s$ for formation of the singularity by letting $t_f=t_s$ (the time when the spherical shell $R_i=r\,f(t_i)$ ends up in the singular shell $R_f=0$). Whereas, in the present study, $t_f\,-\,t_i$ can not be used to
find neither $\Delta\,t_{AH}$ nor $\Delta\,t_s$ since we have already a black hole with an apparent horizon and a hidden singularity at the initial time $t_i$ since we assume that initially
we have a Schwarzschild black hole immersed in the fluid. Eq.(\ref{eq:41a}) tells us that this is really
the case initially, and the size of the horizon gets larger as the gravitational collapse evolves (i.e. as $f$ gets smaller and smaller) and the singularity remains hidden behind the horizon.
In this case $t_{AH}\,-\,t_i$ simply corresponds to the time it takes for a shell $R_i=r\,f(t_i)$ to become the shell of apparent horizon
$R_f=r\,f(t_{AH})=r_1\,f(t)=2M$ (where $r_1$ is given by (\ref{eq:41a})). We give the derivation of $t_{AH}\,-\,t_i$ and $t_s\,-\,t_i$ for this metric in Appendix B for the sake of completeness.

 \section{More detailed study of dust}

 To see the implications of the metric specified in (\ref{eq:31a}) and (\ref{eq:31aa}) more clearly and explicitly we restrict our attention to the case where the fluid is initially a dust. We must have $G^1_{\;1}\,\rightarrow\,0$ as $r\,\rightarrow\,\infty$ if the fluid is initially dust since the causality requires that the black hole does not have effect any effect at $r\,\rightarrow\,\infty$. This requirement together with (\ref{i1c}) and the comment about the initial conditions after (\ref{eq:31aa}) require that, if the fluid initially is a uniform dust, then we must have
 \begin{equation}
 f(t)\,=\,a(t)\,\propto\,t^\frac{2}{3}
 \label{i3}
 \end{equation}
 i.e. $a(t)$ does not mimic the scale factor of dust at initial times but it has the form of the cosmological scale factor of dust for all times. By using Mathematica, the corresponding $G^0_{\;1}$ is found to be
 \begin{eqnarray}
 G^0_{\;1}&=&
-A\,
 \left[4 c^3 r^2 t^{4/3} (3 M-2 r)+6 c^2 r^3 t^{2/3}+9 c t^2 (r-2 M)+9 t^{4/3} (2 M-r)\right]
         \label{G01}
 \end{eqnarray}
 with
 \begin{equation}
 A\,=\,\frac{24 M r}{c}\left(4 c^2 r^4-9 t^{2/3} (r-2 M)^2\right)^{-2}t^{-\frac{5}{3}} \label{G01a}
 \end{equation}
 that have been obtained by using  $f(t)\,=\,a(t)\,=\,c\,t^\frac{2}{3}$ where $c$ is some constant.

 It is evident that the positivity or negativity of $G^0_{\;1}$ depends on the relative magnitudes of $c$, $t$ and $r$. To see the situation better one may write $a(t)\,=\,c\,t_0^\frac{2}{3}\gamma^\frac{2}{3}$ where $\gamma= \frac{t}{t_0}$ and $t_0$ denotes the present time. Then, the common convention $a(t_0)=1$ implies $c=1$. Therefore, (\ref{G01}) and (\ref{G01a}) in this unit system depend only on $t$ and $r$ and we may set $c=1$ in these equations i.e. (\ref{G01}) and (\ref{G01a}) in this unit system reduce to
  \begin{eqnarray}
 G^0_{\;1}&=&
-A\,
 \left[4 r^2 t^{4/3} (3 M-2 r)+6 r^3 t^{2/3}+9 t^2 (r-2 M)+9 t^{4/3} (2 M-r)\right]
         \label{G01b}
 \end{eqnarray}
 with
 \begin{equation}
 A\,=\,24 M r \left(4 r^4-9 t^{2/3} (r-2 M)^2\right)^{-2}t^{-\frac{5}{3}} \label{G01ab}
 \end{equation}
 In the above equations $t=1$ corresponds to present time i.e. to the age of the universe i.e. to the size of the observable universe. Therefore, $r$ that shows the distance from the black hole is characteristically much smaller than $t_0$ (in geometric units). (However, to emphasize the effect of the black hole on the collapse we take $t$ and $r$ to be in the same order of magnitudes in the plots). It is evident from (\ref{G01b}) and (\ref{G01ab}) that depending on the relatives magnitudes of $r$ and $t$, $G^0_{\;1}$ may be either positive or negative. For example, for $r\,\gg\,M$, (\ref{G01b}) becomes
 $-A\,
 r\left[-8 r^2 t^{4/3}  +6 r^2 t^{2/3}+9 t^2 - 9 t^{4/3}\right]$ which is positive for $t\simeq\,t_0=1$, negative for  $t\,\gg\,t_0$ while for small $t$ it may be positive or negative depending on how small $t$ is. In the case of small $r$ with $r\,>\,2M$, $G^0_{\;1}$ may be either positive or negative depending on values of the parameters. For example, for $r=3M$ and $t\sim\,t_0$, $G^0_{\;1}$ is negative. This, in the light of (\ref{eq:1ba}) implies that the effect of accreation may induce a positive or negative pressure in the radial direction. In fact, the plots of $G^0_{\;1}$ and $G^1_{\;1}$ drawn by Mathematica shows
 that this is really the case e.g. as given in Figure \ref{fig:1}. It seems that there is no simple relation between $G^0_{\;1}$ and $G^2_{\;2}=G^3_{\;3}$. In fact, the plots of $G^1_{\;1}$ and $G^2_{\;2}$ imply that both may be of the same sign or may be of different signs e.g. as given in Figures \ref{fig:2} and \ref{fig:3}. $G^0_{\;0}$ in plots turns out to be negative for the values of $r$ outside horizon as should be since it corresponds to $-\rho$. Note that one must be careful about the location of horizon since it increases by time as may be seen from Eq.(\ref{eq:41a}) for $f(t)\,=\,a(t)$, and $G^0_{\;0}$ should be negative for the values of $r$ outside the horizon. We have checked this all graphs we have plotted. It turns out that most of the interesting parameter space corresponds to the regions outside the horizon. For example, for Figure \ref{fig:1}, the situation is shown in Figure \ref{fig:4}.

It is worthwhile to comment on some nontrivial and interesting points about the Einstein tensor and the corresponding energy-momentum tensor of this system. The first
interesting point is the emergence of a radial pressure (as is evident from the non-vanishing $G^1_{\;1}$) for some values of parameters although initially the fluid is a homogeneous isotropic dust, so has no pressure.
As we have mentioned above, from a technical point of view, this is due to emergence of a non-vanishing $G^0_{\;1}$ after the start of the collapse as is evident from (\ref{eq:1ba}). In more physical terms this situation
may be described as follows.
The induced pressure after the start of the collapse is due to the initial condition just after the emergence of the black hole. A dust particle just after the
emergence of the black hole behaves like a test particle in the presence of a Schwarzschild black hole provided that the density of the dust is not extremely large (i.e. provided that we are not at very early times). The Lagrangian of the test particle may be taken
as $m\,\sqrt{g_{\mu\nu}\frac{dx^\mu}{ds}\frac{dx^\nu}{ds}}$.
The corresponding Lagrange equations for the Schwarzschild metric results in a conserved quantity, namely, the total mechanical energy of the particle $\tilde{E}$ which is given by $\tilde{E}= m\,e^{2\lambda} \frac{dt}{ds}$
(i.e. $\frac{dt}{ds}=\frac{\tilde{E}}{m\,e^{2\lambda}}$)
as mentioned after (\ref{eq:8}), so $e^{2\lambda}\left(\frac{dt}{d\tau}\right)^2=\frac{\tilde{E}^2}{m\,e^{2\lambda}}$. Then the identity $g_{\mu\nu}\frac{dx^\mu}{d\tau}\frac{dx^\nu}{d\tau}=-1$
for massive particles at rest results in $\frac{\tilde{E}^2}{m\,e^{2\lambda}}\,=\,1$.
This relation for a dust particle at rest in the presence of a Schwarzschild black hole implies $\tilde{E}\,=\,\sqrt{1-\frac{2M}{r}}$. In other words, the total energy of dust particles in different
spherical shells with
different initial $r$ values have different energies. Although the above analysis
is true for the dust particles at initial times
(that may be identified as test particles), it is plausible to expect the main points of the above argument be applicable at later times too.
Therefore, the shell of particles that are initially at some $r_i$ and arrive the shell at coordinate $r$ at a later times $t$ after the collapse will have higher speeds compared to the ones that arrived earlier.
On the other hand, all particles in a dust seem to be at rest at co-moving frame
i.e. if their relative motion
due to cosmic expansion (i.e. Hubble flow) is subtracted. In other words, the collapsing fluid in the present study does not behave like dust any longer after the start of the collapse.
In this system, if the frame is located at one of the particles then the particles at shells with smaller $r$ seem to move slower while those with
larger $r$ seem to move faster i.e. the particles have non-vanishing kinetic energies (even when Hubble flow is subtracted). This, in turn, implies that the fluid in this system gains some pressure due to
collapse \cite{Weinberg}.
Moreover, this pressure is expected to be negative since the points on a shell with smaller $r$ have smaller kinetic energies unless we consider very late times where all particles that are at a finite distance from the black hole may be considered to be coming from infinity, so with $\tilde{E}=1$ i.e. with the same kinetic energies i.e. with a positive pressure. A similar argument accounts for the non-vanishing of $G^0_{\;1}$ and $G^2_{\;2}=G^3_{\;3}$. The fact
that the speeds of the particles at smaller $r$ values on a shell are smaller implies a non-vanishing outward power flux i.e. $G^0_{\;1}=8\pi\,T^0_{\;1}\neq\,0$. On a point at the mid of the shell,
the particles with larger and smaller $r$ seem to
approach each other in the directions perpendicular to the radial direction i.e. generically negative $G^2_{\;2}=G^3_{\;3}$. This, in turn, results in a pressure in the directions perpendicular to the radial direction. In other words, the effect of the black hole gradually turns
the homogeneous fluid to an inhomogeneous one which develops pressures and accretion through gravitational collapse. Although one may have a rough picture of collapse by the above argument it is insufficient to give the full picture because of the complexity of the system. A more detailed study of the system is needed in future studies.

\section{conclusion}

In this study we have considered gravitational collapse of a fluid (that is initially dust) under the effect of a Schwarzschild black hole that suddenly forms inside the dust.  We have derived a family of metrics after using some assumptions. We have found that this metric corresponds to gravitational collapse of a fluid that includes dust collapse as its subcase although we have started from a homogeneous isotropic dust for the derivation of the metric. To see the phenomenological implications of this metric we have focused on the case of gravitational collapse of a dust. First we have obtained more general analytical conclusions, and then used Mathematica to observe the evolution of the system in the cases where the analytical treatment is insufficient.  We have found interesting results. We have found that the dust in this framework develops radial and tangential pressures that may be positive or negative depending on the parameters. This seems to suggest a local dark energy behaviour for parameter choices.
Although the equation of state parameter in this case would be similar to that of dark energy (which manifests at cosmological scales), this does not automatically imply that such a system would induce an
accelerated expansion of the local universe since the metric here is not the Robertson-Walker metric. However, it would be interesting to see in future studies if this dark energy-like behaviour essentially becomes similar to the cosmological dark energy behaviour if such gravitationally collapsing dust around black hole systems are abundant at cosmological level. We have also discussed
some essential properties
of the corresponding metric such as its singularity structure and its apparent horizons.
We have found that the size of the apparent horizon of this black hole increases. This together with non-vanishing inward radial power flux (shown by the non-vanishing $G^0_{\;1}$) implies that the black hole is accrediting.
Another point worth to mention is that, one may wonder if this metric is another form of Vaidya metric. The massiveness of the dust particles seem to exclude this possibility since accretion or loss of
the black hole is due to null dust for Vaidya metric. This may be expressed in a more general context as follows: The form of the energy momentum tensor for
Vaidya metric has the form $T_{\mu\nu}\,\propto\,\tilde{U}_\mu\tilde{U}_\nu$ where $\tilde{U}_\mu$ is a null 4-vector while the energy-momentum tensor in this case (that is given by (\ref{eq:28r})) has a wholly different form.
In fact, a comparison of the Vaidya's metric with (\ref{eq:31a}) also seem to suggest that this is a wholly new metric
(see Appendix C). We have also checked (in Appendix C) if this metric is another form of McVittie metric. It seems that answer to this question too is negative. Another important point to be addressed is the observational features of this system. One such particular feature would be the
the shadow of the black hole in the present study. In principle this may be done, for example, as in the case where the static limit of the metric is Schwarzschild metric that corresponds to a black hole surrounded by
a shell of dark matter as done in \cite{Konoplya}. However, the situation in this case is more complicated.  This point may be a rather non-trivial project to be addressed in future in a separate study. All these points need further study by their own.

\begin{acknowledgments}
We would like to thank Professor Vitor Cardoso for reading the draft version of the manuscript and for his valuable comments. We would also like to thank the anonymous referee with a special emphasis for pointing out some errors and for his/her detailed, constructive and comprehensive comments and suggestions.

This paper is financially supported by {\it The Scientific and Technical Research Council of Turkey (T{\"{U}}BITAK)} under the project 117F296 in the context of the COST action {\it CA 16104 "GWverse"}
\end{acknowledgments}

\appendix

\section{Transformation of (\ref{eq:1}) into (\ref{eq:43})}
By using the identity
\begin{equation}
dr=\frac{dR-\dot{R}dt}{R^{\prime}},
\end{equation}
where $R=R\left(r,t\right)$, the most general spherically symmetric line element given in the Eq.(\ref{eq:1}) can be rewritten as
\begin{equation}
ds^2=-\bigg(e^{2\lambda}-\frac{\dot{R}^2}{R^{\prime 2}}e^{2\psi}\bigg)dt^2+\frac{e^{2\psi}}{R^{\prime 2}}dR^2-\frac{2\dot{R}e^{2\psi}}{R^{\prime 2}}dtdR+R^2(d\theta^2+\sin^2\theta d\phi^2).\label{eq:A2}
\end{equation}
To get rid of the cross-term $dtdR$, we introduce a new time coordinate $T=T(t,R)$ with $dT=\frac{1}{W}(dt+\beta dR)$, which is an exact differential, so it satisfies $\frac{\partial}{\partial R}(\frac{1}{W})=\frac{\partial}{\partial t}(\frac{\beta}{W})$, where $W(t,R)$ is an integration factor and $\beta (t,R)$ is a function to be set. By defining $\beta(t,R)=\frac{\dot{R}e^{2\psi}}{R^{\prime 2}(e^{2\lambda}-e^{2\psi}\dot{R}^2/R^{\prime 2})}$ in $dT$, the Eq.(\ref{eq:A2}) becomes
\begin{equation}
ds^2=-\bigg(e^{2\lambda}-\frac{\dot{R}^2}{R^{\prime 2}}e^{2\psi}\bigg)W^2dT^2+\frac{e^{2\psi+2\lambda}}{R^{\prime 2}e^{2\lambda}-\dot{R}^2 e^{2\psi}}dR^2+R^2(d\theta^2+\sin^2\theta d\phi^2).
\end{equation}
By using $U$ and $\Gamma$ introduced in the Eq.(\ref{eq:44}), we obtain
\begin{equation}
ds^2=-e^{2\lambda}\left(1-\frac{U^2}{\Gamma^2}\right)W^2dT^2 +\frac{dR^2}{\Gamma^2-U^2}dR^2
+R^2(d\theta^2+\sin^2\theta d\phi^2),
\end{equation}
which is Eq.(\ref{eq:43}).

\section{Derivation of $t_s\,-\,t_{AH}$ for this metric}

In a general analysis that holds for a broad family of metrics (rather than
a particular metric) one may re-express $F(r,t)=R(1-e^{-2\psi}R'^2+e^{-2\lambda}\dot R^2)$ as $\dot{R}^2\,=\,\frac{F}{R}\,+\,s$ where $s$ is a function of $r$ and $t$
in general (while in the case of dust may be taken to a constant), and determine the behaviour of $\frac{F}{R}\,+\,s$ that is expressed in terms of $R=r\,f(t)$ and $t$ in region of space-time, and this equation is used to express
the time $t$ for formation of the apparent horizon and the singularity.

On the other hand, in this tsudy, $F(r,t)=F(r_0)$ is totally fixed up to arbitrary $f(t)$ and $a(t)$. Therefore, if one uses the formula $\dot{R}^2\,=\,\frac{F}{R}\,+\,s$ then
one obtains the trivial equation, $\dot{f}=\dot{f}$. In fact, this tells us that, to obtain a formula for the time of formation of an apparent horizon or a singularity one must specify $\dot{f}$ by
\begin{equation}
\dot{f}(t)\,=\,Y(f),
\label{eq:47}
\end{equation}
 where $Y(f)$ is some function of the scale factor $f(t)$. One may write a similar equation for $a(t)$ (i.e. one may specify if the universe is radiation dominated,
 matter dominated or if it is dominated by some other kind of energy density). However, we will assume that $a(t)$ is already fixed, and focus on the dependence of the singularity and apparent horizon formation times on $f(t)$. A similar analysis may be done for $a(t)$. To obtain concrete results out of (\ref{eq:47}) we must specify $Y(f)$. For example, to have a general on how this works, we let
 \begin{equation}
Y(f)\,=\,(f-f_0)^\gamma,
\label{eq:48}
\end{equation}
 where $f_0$ is some constant. We take $f_0\,>\,f_i$ where $f_i=f(t_i)$ is the initial value of $f$ since (although being small) $\dot{f}(t)$ is not zero in general at the time of the start of the collapse due
 to the non-vanishing cosmological expansion.  We take $f(t)\,<\,f_i$ for $t\,>\,t_i$ since we had found that $\dot{f}\,<\,0$ (i.e. $Y\,<\,0$). Without loss of generality we let $f_i=1$.

 The terms in  (\ref{eq:48}) may be rearranged to express the time $t$ since the start of the collapse as
$t\,=\,\int_{t_i}^{t}dt\,=\,\int_1^f\,\frac{df}{(f-f_0)^\gamma}$. At the time of formation of the singularity $t_s$ we have $R=0$ i.e. $f=f_s=0$ while at the time of formation of
the apparent horizon we have $t=t_{AH}$ i.e. $f=f_{AH}$. $\gamma$ are odd integers by the condition of the reality of $Y(f)$ and $f(t)\,<\,f_i$ for $t\,>\,t_i$, and $Y(f)\,<\,0$.

 For $\gamma=1$ the integration results in,
\begin{equation}
t_s\,-\,t_{AH}\,=\,\ln{\left(\frac{f_0}{f_0-f_{AH}}\right)}
\label{eq:49b}
\end{equation}
We see that the singularity is not naked i.e. $t_s\,-\,t_{AH}\,>\,0$ for $\gamma=1$ if $f_0\,>\,f_{AH}$ and $f_0\,>\,|f_0-f_{AH}|$. On the other hand, for $\gamma\neq\,1$
\begin{equation}
t_s\,-\,t_{AH}\,=\,\frac{1}{1-\gamma}\left[\left(-f_0\right)^{1-\gamma}\,-\,\left(f_{AH}-f_0\right)^{1-\gamma}\right]\, \label{eq:49c}
\end{equation}
We see that (for both positive and negative odd integer values of $\gamma$) we need $|\left(f_{AH}-f_0\right)|\,<\,f_0$ to have $t_s\,-\,t_{AH}\,>\,0$.

\section{Comparison of (\ref{eq:31a}) with Vaidya metric and McVittie metric}

\subsection{Comparison of (\ref{eq:31a}) with Vaidya metric}

The Vaidya metric  in its well-known form reads \cite{Vaidya, diagonal-Vaidya}
\begin{equation}
ds^2=-\left(1-\frac{2M(z)}{r}\right)dz^2+2\,dz\,dr+ r^2(d\theta^2+\sin^2\theta d\phi^2), \label{eq:03ap}
\end{equation}
where $z$ is related to the coordinates $t$ and $r$ by switching to Eddington-Finkelstein coordinates. Eq.(\ref{eq:03ap}) describes radial influx (outflux) of massless particles for  $z=-v$ ($z=u$) where $v$, $u$ are the ingoing and outgoing Eddington-Finkelstein coordinates. This metric may be expressed in a diagonal form \cite{Vaidya2, diagonal-Vaidya} as
\begin{equation}
ds^2=-f_0(r,t)\,dt^2\,+\,\left(1-\frac{2\,F(r,t)}{r}\right)^{-1}dr^2\,+r\,^2(d\theta^2+\sin^2\theta d\phi^2), \label{eq:31x}
\end{equation}
where $f_0(r,t)=\frac{\dot{F}^2(r,t)}{f^2(r,t)}\left(1-\frac{2\,F(r,t)}{r}\right)$ with $f(r,t)=F^\prime(r,t)\left(1-\frac{2\,F(r,t)}{r}\right)$. It is evident that the form of (\ref{eq:31x}) is quite different from that of (\ref{eq:31a}).

\subsection{Comparison of (\ref{eq:31x}) with McVittie metric}

McVittie metric reads
%ds^2=-\frac{\left(1-\frac{M(t)}{2r}\right)^2}{\left(1+\frac{M(t)}{2r}\right)^2}dt^2+ a^2\left(t\right)\left(1+\frac{M(t)}{2r}\right)^4\left(dr^2 + r^2(d\theta%^2+\sin^2\theta d\phi^2)
%\end{equation}
\begin{equation}
ds^2=-\frac{\left(1-\mu\right)^2}{\left(1+\mu\right)^2}dt^2+ \tilde{a}^2\left(t\right)\left(1+\mu\right)^4\left(d\rho^2 + \rho^2(d\theta^2+\sin^2\theta d\phi^2)\right), \label{eq:32ap}
\end{equation}
where $\mu= M/2\tilde{a}\left(t\right)\rho$. The McVittie metric reduces to the Schwarzschild
metric in isotropic coordinates for $\tilde{a}\left(t\right)=1$, and it reduces to the Friedmann-Robertson-Walker (FRW) metric for  $M/\rho \rightarrow 0$.
The Schwarzschild metric in isotropic coordinates may be transformed to its original Schwarzschild form by letting $\tilde{a}\left(t\right)=1$, and $r=\rho\left(1+2M/\rho\right)^2$ in (\ref{eq:32ap}).

The metric (\ref{eq:31a}) can not be reduced to Schwarzschild metric since this would need $\dot{f}=0$ which in the case of dust implies $\dot{a}=0$. One can let $\dot{a}=0$ at any time during the evolution of this system because this would imply $n_0=0$ since it is related to Hubble parameter by $\frac{\dot{a}^2}{a^2}= \frac{8\pi}{3}m\,\frac{n_0}{a^3}$.  If (\ref{eq:31a}) were an extension of the the McVittie metric it would reduce to the Schwarzschild metric in its static limit but a static limit this metric does not exist unless the dust is removed from the system.

\newpage
\begin{figure}[h!]
\centerline{\includegraphics[scale=0.6]{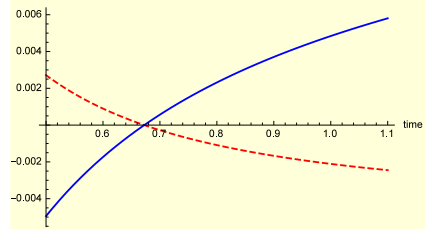}}
\caption{Time versus $G^0_{\;1}$ (blue) and $G^1_{\;1}$ (dashed red) graphs for $c=1$, $M=0.1$, $r =3$}
\label{fig:1}
\end{figure}

\begin{figure}[h!]
\centerline{\includegraphics[scale=0.6]{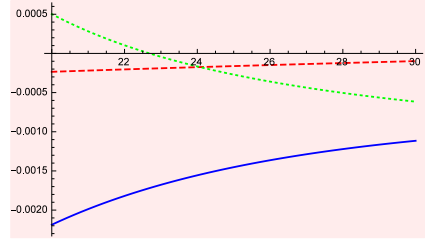}}
\caption{Time versus $G^1_{\;1}$ (dashed red), $G^0_{\;0}$ (blue), $G^2_{\;2}$ (dotted green) graphs for $c=1$, $M=1$, $r =3$}
\label{fig:2}
\end{figure}

\begin{figure}[h!]
\centerline{\includegraphics[scale=0.6]{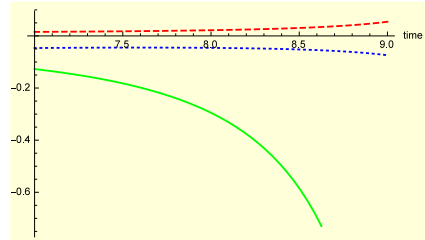}}
\caption{Time versus $G^1_{\;1}$ (dashed red), $G^0_{\;0}$ (dotted blue), $G^2_{\;2}$ (green) graphs for $c=0.1$, $M=1$, $r =30$}
\label{fig:3}
\end{figure}

\begin{figure}[h!]
\centerline{\includegraphics[scale=0.6]{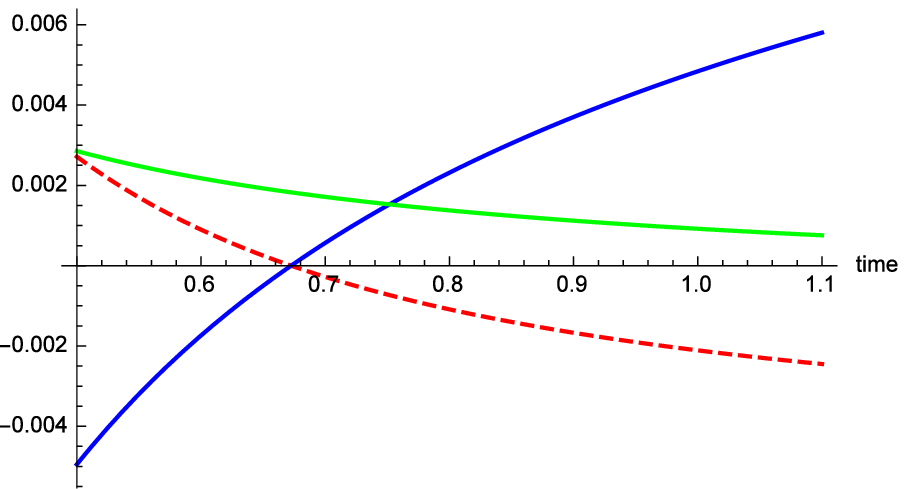}}
\caption{Figure 1 with the function $H=\left(\frac{\dot{a}}{a}\right)^2\,a^3\,r^3-a\,r+2M$ is included. The zeros of $H$ correspond to Eq.(\ref{eq:40}) with $f(t)\,=\,a(t)$ (where $\left(\frac{\dot{a}}{a}\right)^2\,a^3\,=\,\frac{8\pi}{3}m\,n_0$ is used). The green solid line here corresponds to $\frac{H}{10^4}$.}
\label{fig:4}
\end{figure}


\begin{thebibliography}{99}
\bibitem{McVittie}
G.C. McVittie, {\it The Mass-Particle in an Expanding Universe}, {\it Mon. Not. R. Astron. Soc.}, {\bf 93}, 325 (1933).
\bibitem{Tolman}
R.C. Tolman, {\it Effect of Inhomogeneity on Cosmological Models}, {\it Proc. Nat. Acad. Sci.} {\bf 20}, 169 (1934).
\bibitem{Oppenheimer}
J.R. Oppenheimer, H. Snyder, {\it On Continued Gravitational Contraction}, {\it Phys. Rev.} {\bf 56}, 455 (1939).
\bibitem{Bondi}
H. Bondi, {\it Spherically Symmetric Models in General Relativity}, {\it Mon. Not. Roy. Astron. Soc.} {\bf 107}, 410 (1947).
\bibitem{generalized-Bondi}
C. Gao, X. Chen, Y-G. Shen, V. Faraoni, {\it Black holes in the universe: Generalized Lema{\v{i}}tre-Tolman-Bondi solutions}, {\it Phys. Rev. D} {\bf 84}, 104047 (2011), arXiv:1110.6708.
\bibitem{Kaloper}
N. Kaloper, M. Kleban, D. Martin, {\it McVittie's Legacy: Black holes in an expanding universe}, {\it Phys. Rev. D} {\bf 81}, 104044 (2010), arXiv:1003.4777.
%\bibitem{Lemaitre}
%G. Lemaitre, {\it The Expanding Universe}, {\it Annales Soc. Sci. Brux. A} {\bf 53}, 51 (1933) [{\it Gen. Rel. Grav.} {\bf 29}, 641 (1997)]
\bibitem{Vaidya}
P.C. Vaidya, {\it The gravitational field of a radiating star}, {\it Proc. Indian Acad. Sci. A} {\bf 33}, 264 (1951).
\bibitem{Joshi}
P.S. Joshi, D. Malafarina, {\it Recent Developements in Gravitational Collapse and Spacetime Singularities}, {\it Int. J. Mod. Phys.}, {\bf 20}, 2641 (2012), arXiv:1201.3660.
\bibitem{Bambi}
C. Bambi, {\it Black Holes: A Laboratory for Testing Strong Gravity}
(Springer, Singapore, 2017)
\bibitem{types}
P. Martin-Moruno, M. Visser, {\it Essential core of the Hawking-Ellis types}, {\it Class. Quantum Grav.} {\bf 35}, 125003 (2016), arXiv:1802.00865, and the references therein.
\bibitem{Misner} C.W. Misner, D.H. Sharp, {\it Relativistic Equations for Adiabatic, Spherically Symmetric Gravitational Collapse}, {\it Phys. Rev.} {\bf 136} B 571 (1964)
\bibitem{PBH-matter-dominated} T. Kokubu, {\it Effect of inhomogeneity on primordial black hole formation in the matter dominated era}, {\it Phys. Rev. D} {\bf 98}, 123024 (2018), arXiv:1810.03490.
%\bibitem{static-non-static-de-Sitter}  A. Mitra, {\it Interpretational conflicts between the static and non-static forms of the de Sitter metric}, {Sci.Rep.} {\bf 2},   923 (2012)
%\bibitem{Joshi2}
%I.H. Dwivedi, P.S. Joshi, {\it On the Occurence of Naked Singularity in Spherically Symmetric Gravitational Collapse}, {\it Comm. Math. Phys.}, {\bf 166}, 117 (1994).
\bibitem{Joshi3}
T.P. Singh, P.S. Joshi, {\it The final fate of spherical inhomogeneous dust collapse}, {\it Class. Quantum Grav.}, {\bf 13}, 559 (1996), e-print: gr-qc/9409062. \\
S. Jhingan, P.S. Joshi, {\it Structure of Singularity in Spherical Inhomogeneous Dust Collapse}, arXiv: gr-qc/9701016.
\bibitem{Joshi5}
P.S. Joshi, D. Malafarina, {\it All black holes in Lema{\^i}tre-Tolman-Bondi inhomogeneous dust collapse}, {\it Class. Quantum Grav.}, {\bf 32}, 145004 (2015), arXiv:1505.1146.
\bibitem{book}
J. Plebanski, A. Krasinski, {\it An Introduction to General Relativity and Cosmology},
(Cambridge Univ. Press., New York, 2006)
\bibitem{Joshi4}
P.S. Joshi, D. Malafarina, {\it The final fate of spherical inhomogeneous dust collapse}, {\it Class. Quantum Grav.}, {\bf 13}, 559 (1996), e-print: gr-qc/9409062.
\bibitem{Weinberg}
S. Weinberg, {\it Gravitation and Cosmology}, (John Wiley $\&$ Sons, USA, 1972).
\bibitem{galaxies}
V. Faraoni, {\it Evolving Black Hole Horizons in General Relativity and Alternative Gravity}, {\it Glaxies}, {\bf 1}, 114 (2013); e-print: 1309.4915.
\bibitem{Faraoni2}
V. Faraoni, G.F.R. Ellis, J.T. Firouzjaee, A. Helou, I. Musco, {\it Foliation dependence of black hole apparent horizons in spherical symmetry}, {\it Phys. Rev. D} {\bf 95}, 024008 (2017), arXiv:1610.05822.
\bibitem{Konoplya} R.A. Konoplya, {\it Shadow of a black hole surrounded by dark matter}, {\it Phys. Lett. B}, {\bf 795}, 1 (2019), arXiv:1905.00064.
\bibitem{diagonal-Vaidya}
V.A. Berezin, V.I. Dokuchaev, Y.N. Eroshenko, {\it On maximal analytic extension of Vaidya metric}, {\it Class. Quantum Grav.} {\bf 33}, 145003 (2016), arXiv:1603.00849; \\
V.A. Berezin, V.I. Dokuchaev, Y.N. Eroshenko, {\it Vaidya Spacetime in the Diagonal Coordinates}, {\it J. Exp. Theor. Phys.} {\bf 124}, 446 (2017), arXiv:1704.06889.
\bibitem{Vaidya2}
P.C. Vaidya, {\it The External Field of a Radiating Star in General Relativity}, {\it Current Science} {\bf 12}, 183 (1943) ({\it Gen. Relativ. Gravit} {\bf 31}, 119 (1999)).




\end{thebibliography}
\end{document}